\documentclass[12pt,a4paper]{article}

\usepackage{authblk}
\usepackage[T1]{fontenc}
\usepackage[utf8]{inputenc}
\usepackage[protrusion=true,expansion=false]{microtype}
\usepackage[margin=1in]{geometry}
\usepackage{float}
\usepackage{hyperref}
\usepackage{doi}

\usepackage{amsmath,amssymb,amsthm,mathtools}
\usepackage{bm}

\usepackage{booktabs}
\usepackage{array}
\usepackage{caption}
\usepackage{graphicx}

\hypersetup{
    colorlinks=true,
    linkcolor=blue,
    citecolor=blue,
    urlcolor=blue
}

\newcommand{\Ahat}{\hat{A}}
\newcommand{\gcm}{\,\text{g\,cm}^{-3}}
\newcommand{\GeV}{\,\text{GeV}}
\newcommand{\eV}{\,\text{eV}}
\newcommand{\km}{\,\text{km}}
\newcommand{\Hf}{H_{\mathrm{f}}}
\newcommand{\Htf}{\tilde{H}_{\mathrm{f}}}
\DeclareMathOperator{\Tr}{Tr}
\DeclareMathOperator{\diag}{diag}
\DeclareMathOperator{\Adj}{Adj}
\newcommand{\su}{\mathfrak{su}}
\newcommand{\un}{\mathfrak{u}}
\newcommand{\JCP}{J_{\mathrm{CP}}}
\newcommand{\Jmat}{J_{\mathrm{mat}}}

\newcommand{\Lmagic}{L_{\mathrm{magic}}}
\newcommand{\ket}[1]{\left|#1\right\rangle}
\newcommand{\bra}[1]{\left\langle#1\right|}

\title{\textbf{Algebraic Structure of Three-Flavor Neutrino Oscillations in Constant-Density Matter: Cayley--Hamilton Evolution, DMP Resummation, and Closed-Form Uncertainty Propagation}}

\author[1]{Aaryan Chaulagain}
\author[1]{Anju Dhakal}
\author[1]{Daya Nidhi Chhatkuli\thanks{Corresponding author:
  \href{mailto:chhatkulidn@gmail.com}{chhatkulidn@gmail.com}}}
\affil[1]{Department of Physics, Tri-Chandra Multiple Campus,
         Tribhuvan University, Nepal}
\date{}

\begin{document}
\maketitle

\begin{abstract}
For three-flavor neutrino oscillations in constant-density matter, the
Cayley--Hamilton theorem forces the evolution operator into a quadratic
polynomial in $\Hf$, with coefficients determined by the three real
eigenvalues through a Vandermonde system we write out explicitly. The
eigenvalues follow from Cardano's trigonometric formula, recovering the
Zaglauer--Schwarzer expressions. The Denton--Minakata--Parke (DMP)
approximation achieves fractional accuracy better than $10^{-4}$ because
its $1$--$3$ rotation is a resummation: it removes the near-degeneracy
that makes the naive expansion diverge at $\Ahat\to1$, replacing the
unbounded $(1-\Ahat)^{-1}$ with an effective parameter
$\epsilon_0\lesssim0.015$ bounded uniformly in energy. A density-matrix
treatment with a Lindblad term handles open-system decoherence and
wave-packet effects in the same language; matter-dressed coherence
lengths satisfy $L/L^{ij}_{\mathrm{coh}}\sim10^{-3}$--$10^{-2}$ for
terrestrial baselines. The CP asymmetry
$\mathcal{A}_{\mathrm{CP}}(\nu_\mu\to\nu_e)$ is split into genuine and
matter-induced fake contributions. Closed-form Jacobians in the
NuFIT~6.0 parameter basis feed Monte Carlo and linearized
uncertainty-propagation schemes, the latter validated against a
Feldman--Cousins profile-likelihood mapping near physical boundaries.
The Denton--Parke NuFast-LBL algorithm [Phys.\ Rev.\ D \textbf{110},
073005 (2024)] remains the tool of choice for production fits; the
analytic expressions here supply what iterative solvers cannot---parameter
continuity, transparent limits, and Jacobians in closed form.
\end{abstract}

\textbf{Keywords:} Neutrino oscillations, MSW effect, Perturbative
resummation, $\su(3)$ structure, Density matrix, Wave-packet
decoherence, CP violation, Uncertainty propagation, PMNS matrix.

\newpage
\tableofcontents
\newpage

\section{Introduction}
\label{sec:intro}

Neutrino flavor oscillation is now a precision measurement problem. The
three mixing angles, two mass-squared differences, and the CP-violating
phase $\delta_{\mathrm{CP}}$ of the
Pontecorvo--Maki--Nakagawa--Sakata (PMNS) matrix~\cite{Pont57,Pont58,MNS1962}
have been determined by reactor and accelerator experiments over the past
two decades---Daya Bay~\cite{DayaBay2012}, RENO~\cite{RENO2012}, Double
Chooz~\cite{DoubleChooz2012}, T2K~\cite{T2K2011}, and NOvA~\cite{NOvA2016},
building on the discovery of atmospheric and solar flavor conversion at
Super-Kamiokande~\cite{SuperK1998} and SNO~\cite{SNO2002} and the reactor
confirmation by KamLAND~\cite{KamLAND2003}. The next generation---DUNE~\cite{DUNE2021},
Hyper-Kamiokande (HK)~\cite{HK2018}, and JUNO~\cite{JUNO2022}---aims to
measure leptonic CP violation at the five-sigma level and to resolve the
mass ordering and the $\theta_{23}$ octant. At that precision the
oscillation probability calculation is no longer an afterthought: for
DUNE, matter effects from coherent forward
scattering~\cite{Wolfenstein1978,MSW1985} contribute $10$--$20\%$
corrections to $P_{\mu e}$, and a systematic error in treating them maps
directly onto the inferred value of $\delta_{\mathrm{CP}}$~\cite{Barger1980,Kelly2018}.

The tools for computing those corrections have matured considerably.
Zaglauer and Schwarzer~\cite{ZaglauerSchwarz1988} gave the closed-form
cubic eigenvalues decades ago. More recently the Denton--Minakata--Parke
(DMP) approximation~\cite{DMP2016} reached fractional accuracy below
$10^{-4}$ with a single rotation, and the NuFast-LBL algorithm of Denton
and Parke~\cite{Denton2024} brought the cost down to about
$0.1\,\mu\mathrm{s}$ per evaluation, roughly fifty times faster than a
full eigensystem solve. Modern analyses use
Feldman--Cousins~\cite{Feldman1998} or nested-sampling methods requiring
$\mathcal{O}(10^6)$--$\mathcal{O}(10^7)$ evaluations per fit, which has
driven the sustained effort to produce fast, accurate analytic
expressions catalogued by Barenboim, Denton, Parke, and
Ternes~\cite{LookingGlass2019}. A contemporaneous pedagogical overview
of the three-flavor paradigm is given by Denton~\cite{Denton2025}; the
present work differs in emphasizing the algebraic structure of the
evolution operator, the resummation interpretation of the DMP rotation,
and the closed-form Jacobian infrastructure for uncertainty propagation.

What is missing is not any single ingredient but a treatment that ties
them together for a reader who already knows the field. Take the
evolution operator. That $S(L)$ is quadratic in $\Hf$ follows at once
from Cayley--Hamilton, yet the neutrino literature tends to leave this
implicit, and the Vandermonde system that turns the three eigenvalues
into the three polynomial coefficients is, as far as we have found,
nowhere written out. The DMP rotation is another case. Everyone knows it
tames the $\Ahat\to1$ divergence; less often noted is \emph{why}, namely
that it acts as a resummation and produces an effective expansion
parameter $\epsilon_0$ that remains bounded at every energy. And the
closed-form Jacobians one needs to push parameter uncertainties through
the probabilities---tied to the NuFIT~6.0 covariance and cross-checked
with Feldman--Cousins near the boundaries---we have not seen collected in
one place. These are the threads this paper pulls together, along the way
giving a common density-matrix language for the wave-packet and
open-system corrections.

The organization follows the physics. Sections~\ref{sec:notation}
and~\ref{sec:hamiltonian} fix conventions and build the matter
Hamiltonian; Sections~\ref{sec:su3} and~\ref{sec:diag} develop its
group-theoretic and spectral structure, including the adjugate
eigenvectors and the convergence of iterative refinement. The
density-matrix formalism (Section~\ref{sec:density}) carries through to
the perturbative expansion and its $\Ahat=1$ divergence
(Section~\ref{sec:perturb}), the wave-packet corrections
(Section~\ref{sec:wavepacket}), the MSW resonance and the sub-resonant
T2K/HK regime (Section~\ref{sec:msw}), and the resummation reading of DMP
(Section~\ref{sec:resum}). The remaining sections cover CP and T
asymmetries (Section~\ref{sec:cp}), the magic baseline
(Section~\ref{sec:magic}), uncertainty propagation
(Section~\ref{sec:unc}), analytic limiting-case checks
(Section~\ref{sec:valid}), and extensions (Section~\ref{sec:extensions}),
before we conclude.

\section{Notation and Physical Inputs}
\label{sec:notation}

\subsection{PMNS Parameterization}
\label{sec:pmns}
We adopt the Particle Data Group (PDG) conventions~\cite{PDG2022}. Flavor
eigenstates $\ket{\nu_\alpha}$, $\alpha\in\{e,\mu,\tau\}$, and mass
eigenstates $\ket{\nu_i}$, $i\in\{1,2,3\}$, are related by the unitary
PMNS matrix $U$:
\begin{equation}
  \ket{\nu_\alpha}=\sum_{i=1}^{3}U_{\alpha i}\ket{\nu_i},
  \qquad UU^\dagger=U^\dagger U=\mathbf{1}.
  \label{eq:pmns}
\end{equation}
The standard decomposition is
$U=R_{23}\,\Gamma_\delta\,R_{13}\,\Gamma_\delta^\dagger\,R_{12}$,
with $R_{ij}$ a real rotation by $\theta_{ij}$ and
$\Gamma_\delta=\diag(1,1,e^{i\delta_{\mathrm{CP}}})$. Explicitly, with
$c_{ij}\equiv\cos\theta_{ij}$ and $s_{ij}\equiv\sin\theta_{ij}$,
\begin{equation}
  U=
  \begin{pmatrix}
    c_{12}c_{13} & s_{12}c_{13} & s_{13}e^{-i\delta_{\mathrm{CP}}} \\
    -s_{12}c_{23}-c_{12}s_{23}s_{13}e^{i\delta_{\mathrm{CP}}} &
     c_{12}c_{23}-s_{12}s_{23}s_{13}e^{i\delta_{\mathrm{CP}}} &
     s_{23}c_{13} \\
     s_{12}s_{23}-c_{12}c_{23}s_{13}e^{i\delta_{\mathrm{CP}}} &
    -c_{12}s_{23}-s_{12}c_{23}s_{13}e^{i\delta_{\mathrm{CP}}} &
     c_{23}c_{13}
  \end{pmatrix}.
  \label{eq:umat}
\end{equation}
Majorana phases do not affect oscillation
probabilities~\cite{Bilenky1987} and are omitted. The mass-squared
differences are $\Delta m^2_{21}\equiv m^2_2-m^2_1$,
$\Delta m^2_{31}\equiv m^2_3-m^2_1$, and
$\Delta m^2_{32}=\Delta m^2_{31}-\Delta m^2_{21}$. Normal ordering (NO)
has $\Delta m^2_{31}>0$. A useful combination is the electron-weighted
splitting~\cite{Nunokawa2005,Parke2016}
\begin{equation}
  \Delta m^2_{ee}\equiv
  c^2_{12}\Delta m^2_{31}+s^2_{12}\Delta m^2_{32}
  =\Delta m^2_{31}-s^2_{12}\Delta m^2_{21},
  \label{eq:dmee}
\end{equation}
which appears in $\nu_e$ disappearance and in the DMP
scheme~\cite{LookingGlass2019,DMP2016}.

\subsection{NuFIT 6.0 Parameter Inputs}
\label{sec:nufit}
Central values are from the NuFIT~6.0 global analysis~\cite{NuFIT6}
(Table~\ref{tab:nufit}).
\begin{table}[ht]
\centering
\caption{NuFIT~6.0 best-fit oscillation parameters and $1\sigma$ ranges,
from the global fit including IceCube and Super-Kamiokande atmospheric
data (the IC24 with-SK column of~\cite{NuFIT6,NuFITweb}). For normal
ordering (NO) this is the global best fit; the inverted-ordering (IO)
column is disfavored by $\Delta\chi^2=6.1$. Derived quantities
($\alpha$, $\epsilon_{13}$, $J^{\max}_{\mathrm{CP}}$) are computed from
the central values in this table.}
\label{tab:nufit}
\begin{tabular}{lcc}
\toprule
Parameter & Normal Ordering & Inverted Ordering \\
\midrule
$\theta_{12}$ [$^\circ$] & $33.68^{+0.73}_{-0.70}$ & $33.68^{+0.73}_{-0.70}$ \\
$\theta_{13}$ [$^\circ$] & $8.56^{+0.11}_{-0.11}$ & $8.59^{+0.11}_{-0.11}$ \\
$\theta_{23}$ [$^\circ$] & $43.3^{+1.0}_{-0.8}$ & $47.9^{+0.7}_{-0.9}$ \\
$\delta_{\mathrm{CP}}$ [$^\circ$] & $212^{+26}_{-41}$ & $274^{+22}_{-25}$ \\
$\Delta m^2_{21}$ [$10^{-5}\eV^2$] & $7.49^{+0.19}_{-0.19}$ & $7.49^{+0.19}_{-0.19}$ \\
$\Delta m^2_{3\ell}$ [$10^{-3}\eV^2$] & $+2.534^{+0.025}_{-0.023}$ & $-2.510^{+0.024}_{-0.025}$ \\
$\sin^2\theta_{13}$ & $0.02215$ & $0.02231$ \\
$\sin^2\theta_{23}$ & $0.470$ & $0.550$ \\
$\alpha\equiv\Delta m^2_{21}/|\Delta m^2_{3\ell}|$ & $0.0296$ & $0.0298$ \\
$\epsilon_{13}\equiv\sin\theta_{13}$ & $0.149$ & $0.149$ \\
$J^{\max}_{\mathrm{CP}}$ [$\times10^{-2}$] & $3.35$ & $3.36$ \\
\bottomrule
\end{tabular}
\end{table}
The two small parameters are
\begin{equation}
  \alpha\equiv\frac{\Delta m^2_{21}}{\Delta m^2_{31}}\approx0.0296,
  \qquad
  \epsilon_{13}\equiv\sin\theta_{13}\approx0.149.
  \label{eq:smallparams}
\end{equation}
Since $\epsilon^2_{13}\approx0.022\approx\mathcal{O}(\alpha)$, the
unified counting $\alpha\sim\epsilon^2_{13}$ applies throughout. We note
that the NO global best fit now places $\theta_{23}$ in the first octant
($\sin^2\theta_{23}=0.470$) and gives a CP phase
$\delta_{\mathrm{CP}}\approx212^\circ$ that is consistent with CP
conservation at the $1\sigma$ level~\cite{NuFIT6}; the large
$\delta_{\mathrm{CP}}$ uncertainty ($\pm\sim30^\circ$) propagates
directly into the appearance-channel uncertainty budget of
Section~\ref{sec:unc}.

\subsection{Matter Configuration and Experimental Benchmarks}
\label{sec:matter}
For the reference examples we adopt the average crust configuration along
a $295\km$ baseline: $\rho=2.6\gcm$, $Y_e=0.5$, $m_N=0.939\GeV$,
$L=295\km$. Table~\ref{tab:exp} lists the experimental benchmarks. The
Fermi constant is $G_F=1.1663787\times10^{-5}\GeV^{-2}$ and
$1\km=5.06773\times10^{9}\eV^{-1}$.
\begin{table}[ht]
\centering
\caption{Experimental configurations used in illustrative comparisons.
$E_{\mathrm{peak}}$ is the energy of the first oscillation maximum;
densities are column-density averages.}
\label{tab:exp}
\begin{tabular}{lcccc}
\toprule
Experiment & $L$ [km] & $\rho$ [$\mathrm{g/cm}^3$] & $E_{\mathrm{peak}}$ [GeV] & Ref.\\
\midrule
T2K/HK & 295 & 2.6 & 0.60 & \cite{T2K2011,HK2018}\\
NOvA   & 810 & 2.84 & 1.7 & \cite{NOvA2016}\\
DUNE   & 1300 & 2.84 & 2.5 & \cite{DUNE2021,Kelly2018}\\
JUNO   & 52.5 & 2.6 & 0.004 & \cite{JUNO2022}\\
\bottomrule
\end{tabular}
\end{table}

\paragraph{DUNE baseline geology.}
The $1300\km$ DUNE chord from Fermilab to the Sanford Underground
Research Facility reaches a maximum depth of $\sim32\km$, within the
upper continental crust and never approaching the mantle. Kelly and
Parke~\cite{Kelly2018} showed that a path-averaged constant density
$\rho\approx2.84\gcm$ reproduces the exact oscillation probability at the
$\mathcal{O}(1\%)$ level for both the shape and normalization of the
profile along this chord, so that a single-layer constant-density
treatment suffices and PREM profile layering is not required for DUNE. We
adopt this throughout and do not repeat the argument elsewhere.

\section{Effective Hamiltonian in Matter}
\label{sec:hamiltonian}

\subsection{Derivation from the Electroweak Lagrangian}
\label{sec:lagrangian}
Neutrinos propagating through ordinary matter experience an
index-of-refraction effect from coherent forward scattering off ambient
fermions. The charged-current contribution from $\nu_e$--$e^-$ exchange,
in the zero-momentum-transfer limit, collapses to a local four-fermion
operator; the coherent sum over the electron background yields the
effective potential~\cite{Wolfenstein1978}
\begin{equation}
  V_{\mathrm{CC}}=\sqrt{2}\,G_F N_e
  =\sqrt{2}\,G_F\,\frac{\rho Y_e}{m_N}.
  \label{eq:vcc}
\end{equation}
For the reference configuration $V_{\mathrm{CC}}\approx9.8\times10^{-14}\eV$.
The neutral-current amplitude is flavor-universal and proportional to the
identity in flavor space; it drops out after a global phase
redefinition.

\subsection{Flavor-Basis Hamiltonian and Schr\"odinger Equation}
\label{sec:schrodinger}
The matter-modified flavor Hamiltonian is
\begin{equation}
  \Hf=\frac{1}{2E}\,U\,\diag(0,\Delta m^2_{21},\Delta m^2_{31})\,U^\dagger
      +\diag(V_{\mathrm{CC}},0,0),
  \label{eq:hf}
\end{equation}
after subtracting $m^2_1/(2E)\cdot\mathbf{1}$. $\Hf$ is Hermitian, and
the flavor state satisfies
$i\,d\ket{\nu(x)}/dx=\Hf\ket{\nu(x)}$. For constant density,
\begin{equation}
  \ket{\nu(L)}=S(L)\ket{\nu(0)},
  \qquad S(L)=\exp(-i\Hf L),
  \label{eq:evop}
\end{equation}
with $S(L)$ unitary, so $\sum_\beta P_{\alpha\beta}(L)=1$.

\subsection{Numerical Stability: Hermitian Symmetrization}
\label{sec:symm}
In floating-point arithmetic, construction of $U$ introduces
anti-Hermitian noise at $\mathcal{O}(10^{-16})$. Projecting onto the
Hermitian subspace via
$\Hf\leftarrow\tfrac{1}{2}(\Hf+\Hf^\dagger)$ removes this noise without
altering the physics; afterward unitarity violations in $S(L)$ are
suppressed below $\mathcal{O}(10^{-10})$.

\section{SU(3) Structure and the Evolution Operator}
\label{sec:su3}

\subsection{Embedding in the Lie Algebra \texorpdfstring{$\un(3)$}{u(3)}}
\label{sec:liealg}
$\Hf$ is an element of $\un(3)=\un(1)\oplus\su(3)$, with the $\un(1)$
factor the trace and the $\su(3)$ factor the traceless part. Since the
trace contributes only an overall phase to $S(L)$,
\begin{equation}
  \Hf=\frac{\Tr\Hf}{3}\,\mathbf{1}+\Htf,
  \qquad \Tr\Htf=0,\quad\Htf\in\su(3).
  \label{eq:u3split}
\end{equation}
The Gell-Mann matrices $\{\lambda_a\}^8_{a=1}$~\cite{GellMann1962} give
$\Htf=\sum_a h_a\lambda_a/2$ with $h_a=\Tr(\Hf\lambda_a)$
(Appendix~\ref{app:gellmann}).

\subsection{Invariants and the Characteristic Polynomial}
\label{sec:invariants}
With $\Lambda\equiv2E\Hf$ and invariants $I_n=\Tr(\Hf^n)$, the
characteristic equation is
\begin{equation}
  \Lambda^3-I_1\Lambda^2
  +\tfrac{1}{2}(I_1^2-I_2)\Lambda
  -\tfrac{1}{6}(I_1^3-3I_1I_2+2I_3)=0.
  \label{eq:charpoly}
\end{equation}
In terms of $a=2EV_{\mathrm{CC}}$,
\begin{align}
  A&=I_1=\Delta m^2_{21}+\Delta m^2_{31}+a,
  \label{eq:capA}\\
  B&=\tfrac{1}{2}(I_1^2-I_2)=\Delta m^2_{21}\Delta m^2_{31}
     +a\!\left[\Delta m^2_{21}(1-|U_{e2}|^2)
              +\Delta m^2_{31}(1-|U_{e3}|^2)\right],
  \label{eq:capB}\\
  C&=\det(2E\Hf)=a\,\Delta m^2_{21}\Delta m^2_{31}|U_{e1}|^2.
  \label{eq:capC}
\end{align}
These are the Le~Verrier--Faddeev coefficients~\cite{Abdullahi2022},
computable without explicit matrix inversion.

\subsection{Cayley--Hamilton Structure of the Evolution Operator}
\label{sec:cayley}
By the Cayley--Hamilton theorem, $\Hf$ satisfies its own characteristic
polynomial, so any analytic function $f(\Hf)$ is a polynomial of degree
at most two,
\begin{equation}
  f(\Hf)=c_0\mathbf{1}+c_1\Hf+c_2\Hf^2,
  \label{eq:chpoly}
\end{equation}
with $f(\omega_i)=c_0+c_1\omega_i+c_2\omega_i^2$ for the three eigenvalues
$\omega_i$ of $\Hf$ (energy units). These are related to the rescaled
mass-squared eigenvalues $\lambda_i$ of $\Lambda=2E\Hf$ used from
Section~\ref{sec:diag} onward by $\omega_i=\lambda_i/(2E)$. Applied to
$f(\omega)=e^{-i\omega L}$,
\begin{equation}
  S(L)=c_0(L)\,\mathbf{1}+c_1(L)\,\Hf+c_2(L)\,\Hf^2,
  \label{eq:evoppoly}
\end{equation}
with the coefficients fixed by the Vandermonde system
\begin{equation}
  \begin{pmatrix}
    1 & \omega_1 & \omega_1^2 \\
    1 & \omega_2 & \omega_2^2 \\
    1 & \omega_3 & \omega_3^2
  \end{pmatrix}
  \begin{pmatrix} c_0 \\ c_1 \\ c_2 \end{pmatrix}
  =
  \begin{pmatrix}
    e^{-i\omega_1 L} \\ e^{-i\omega_2 L} \\ e^{-i\omega_3 L}
  \end{pmatrix},
  \qquad \omega_i=\lambda_i/(2E).
  \label{eq:vandermonde}
\end{equation}
The three-flavor amplitude in constant-density matter is thus determined
entirely by the three eigenvalues. While the Cayley--Hamilton property
is implicit in the matrix-exponential literature, the explicit
Vandermonde system~\eqref{eq:vandermonde} connecting eigenvalues to
evolution-operator coefficients is not, to our knowledge, written out in
this form in prior neutrino references.

\section{Exact Hamiltonian Diagonalization}
\label{sec:diag}

\subsection{Exact Cubic Eigenvalue Solution}
\label{sec:cubic}
It is convenient to diagonalize the rescaled matrix $\Lambda\equiv2E\Hf$,
whose coefficients $A$, $B$, $C$ (Eqs.~\ref{eq:capA}--\ref{eq:capC}) carry
dimensions of mass-squared; we denote its eigenvalues $\lambda_i$
(units eV$^2$), so that in vacuum $\lambda_i\to m_i^2-m_1^2$ and the
eigenvalue differences reduce to $\Delta\lambda_{ij}\to\Delta m^2_{ij}$.
The physical phase entering the evolution operator and the oscillation
probabilities is $\lambda_i L/(2E)$; accordingly the oscillation
probabilities of Section~\ref{sec:probgeneral} are written with the
combination $\Delta\lambda_{ij}/(2E)$, which is dimensionally an inverse
length. Substituting $\lambda=\mu+A/3$
gives the depressed cubic
$\mu^3+p\mu+q=0$ with $p=B-A^2/3$ and $q=-2A^3/27+AB/3-C$. Since $\Lambda$
is Hermitian, all roots are real, the discriminant
$\Delta_{\mathrm{cubic}}=-4p^3-27q^2\ge0$, and $p<0$. The three real
roots follow from Cardano's trigonometric
method~\cite{Barger1980,Cardano1545},
\begin{equation}
  \lambda_k=\frac{A}{3}
  +\frac{2}{3}\sqrt{A^2-3B}\;
  \cos\!\left(
    \frac{1}{3}\arccos\!\left(
      \frac{2A^3-9AB+27C}{2(A^2-3B)^{3/2}}
    \right)-\frac{2\pi k}{3}
  \right),\quad k=0,1,2,
  \label{eq:zs}
\end{equation}
the Zaglauer--Schwarzer result~\cite{ZaglauerSchwarz1988}. The cosine
(not hyperbolic) form is the correct branch for a Hermitian matrix with
three real roots.

\subsection{DMP Approximation and Its Accuracy}
\label{sec:dmp}
Defining $x\equiv a/\Delta m^2_{ee}$, DMP~\cite{DMP2016} approximate the
largest eigenvalue by
\begin{equation}
  \lambda_3\approx\Delta m^2_{31}
  +\frac{\Delta m^2_{ee}}{2}\!\left(
    x-1+\sqrt{(1-x)^2+4xs^2_{13}}
  \right).
  \label{eq:dmpapprox}
\end{equation}
The fractional accuracy relative to the exact eigenvalues is better than
$10^{-4}$ for terrestrial energies and both
orderings~\cite{LookingGlass2019,Denton2024}, as in
Table~\ref{tab:accuracy}. A single Newton--Raphson correction,
\begin{equation}
  \lambda_3\to\lambda_3-\frac{X(\lambda_3)}{X'(\lambda_3)},
  \qquad X(\lambda)=\lambda^3-A\lambda^2+B\lambda-C,
  \label{eq:newton}
\end{equation}
improves the fractional precision by several orders of
magnitude~\cite{Denton2024}.

\begin{table}[ht]
\centering
\caption{Fractional eigenvalue/probability accuracy $|\Delta P|/P$ at the
first oscillation maximum for the DUNE configuration
($L=1300\km$, $\rho=2.84\gcm$). Accuracy figures and the cross-scheme
comparison are taken from~\cite{LookingGlass2019,DMP2016,Denton2024}; the
perturbative expressions are reproduced here as theoretical anchors and
were not independently benchmarked for speed. As emphasized
in~\cite{Denton2024}, NuFast-LBL is both faster and more accurate than
the pre-NuFast schemes; speed is therefore not compared in a common
column.}
\label{tab:accuracy}
\begin{tabular}{lll}
\toprule
Expression & Order & $|\Delta P|/P$ \\
\midrule
Madrid~\cite{Cervera2000} & $\mathcal{O}(\epsilon,\alpha)$ & $\sim10^{-2}$ \\
AJLOS~\cite{AJLOS2004}    & $\mathcal{O}(\epsilon,\alpha)$ & $\sim10^{-2}$ \\
Freund~\cite{Freund2001}  & $\mathcal{O}(\epsilon)$        & $\sim10^{-2}$ \\
AKT~\cite{AKT2014}        & $\mathcal{O}(\epsilon,\alpha,a)$ & $\sim10^{-3}$ \\
MP~\cite{MP2016}          & $\mathcal{O}(\alpha)$          & $\sim10^{-2}$ \\
DMP0~\cite{DMP2016}       & $\mathcal{O}(\epsilon_0)$      & $\sim10^{-3}$ \\
DMP1~\cite{DMP2016}       & $\mathcal{O}(\epsilon_0^2)$    & $\sim10^{-5}$ \\
AM5/2~\cite{AsanoMinakata2011} & $\mathcal{O}(s^{5/2}_{13})$ & $\sim10^{-3}$ \\
NuFast(0)~\cite{Denton2024} & exact $+0\,\mathrm{NR}$ & $\sim10^{-4}$ \\
NuFast(1)~\cite{Denton2024} & exact $+1\,\mathrm{NR}$ & $\sim10^{-9}$ \\
Exact (ZS)~\cite{ZaglauerSchwarz1988} & --- & 0 (ref.) \\
\bottomrule
\end{tabular}
\end{table}

\subsection{Eigenvectors via the Adjugate Method}
\label{sec:adjugate}
Once the eigenvalues $\lambda_i$ are known, the squared elements of the
eigenvector matrix $V_e$ follow from the adjugate of $(\lambda I-\Hf)$,
as developed by Abdullahi and Parke~\cite{Abdullahi2022}:
\begin{equation}
  |V_{e\alpha i}|^2=
  \frac{\lambda_i^2-S_{\alpha\alpha}\lambda_i+T_{\alpha\alpha}}
       {\Delta\lambda_{ij}\,\Delta\lambda_{ik}},
  \label{eq:adjugate}
\end{equation}
where $j,k\neq i$ and
\begin{align}
  S_{\alpha\alpha}&=A-(2E)H_{f\alpha\alpha},
  \label{eq:Salpha}\\
  T_{\alpha\alpha}&=(2E)^2[\Adj(\Hf)]_{\alpha\alpha}.
  \label{eq:Talpha}
\end{align}
The numerator quantities $S_{\alpha\alpha}$ and $T_{\alpha\alpha}$ are
the diagonal elements of the adjugate of $(2E\Hf)$; Eq.~\eqref{eq:adjugate}
is thus the adjugate formula of~\cite{Abdullahi2022}. The closely related
eigenvector--eigenvalue identity of Denton, Parke, Tao, and
Zhang~\cite{EigvecIdent2022} expresses the same moduli through products
of eigenvalue differences for sub-matrices; we use the adjugate form
because $\Adj(\Hf)$ is already available from the Le~Verrier--Faddeev
coefficients of Section~\ref{sec:invariants}.

\paragraph{Denominator protection.}
The denominators $\Delta\lambda_{ij}\,\Delta\lambda_{ik}$ become small
near the atmospheric resonance ($\Ahat\to1$). At resonance the minimum
gap is
\begin{equation}
  \Delta\lambda^{\min}_{31}=\frac{\Delta m^2_{31}}{2E}\sin2\theta_{13}.
  \label{eq:gapmin}
\end{equation}
For the crustal benchmark the atmospheric resonance falls at
$E_{\mathrm{res}}\approx12\GeV$ (Section~\ref{sec:msw}); at that energy,
with $\Delta m^2_{31}\approx2.53\times10^{-3}\eV^2$ and
$\sin2\theta_{13}\approx0.29$,
\begin{equation}
  |\Delta\lambda_{31}|_{\min}
  \approx\frac{2.53\times10^{-3}\eV^2}{2\times12\times10^{9}\eV}
         \times0.29
  \approx3\times10^{-14}\eV.
  \label{eq:gapnum}
\end{equation}
The relative error from IEEE~754 double precision is bounded by
\begin{equation}
  \frac{\delta\lambda}{|\Delta\lambda_{ij}|}
  \lesssim\epsilon_{\mathrm{mach}}\cdot\kappa_{\mathrm{Vand}},
  \qquad
  \kappa_{\mathrm{Vand}}\equiv
  \frac{\max_k|\lambda_k|}{|\Delta\lambda_{ij}|_{\min}},
  \label{eq:relerr}
\end{equation}
with $\epsilon_{\mathrm{mach}}\approx2.2\times10^{-16}$. At the operating
energies of all long-baseline experiments (Table~\ref{tab:exp}), which lie
well below the crustal resonance at $E_{\mathrm{res}}\approx12\GeV$, the
denominator $|\Delta\lambda_{31}|$ is comparable to its vacuum value
$\Delta m^2_{31}$ and $\kappa_{\mathrm{Vand}}=\mathcal{O}(1)$; the precision
loss is at the level of $\epsilon_{\mathrm{mach}}$ itself and is negligible
for any foreseeable observable. The concern grows only as $E\to E_{\mathrm{res}}$,
where $\kappa_{\mathrm{Vand}}$ rises; implementations should test
$|\Delta\lambda_{ij}|>\varepsilon_{\mathrm{gap}}$ before invoking
Eq.~\eqref{eq:adjugate}, branching to the DMP approximation
(Eq.~\ref{eq:dmpapprox}) when the criterion fails. The matter Jarlskog
invariant follows from the Naumov--Harrison--Scott
identity~\cite{Naumov1992,HarrisonScott2002},
\begin{equation}
  \Jmat=J\prod_{i>j}\frac{\Delta m^2_{ij}}{\Delta\lambda_{ij}},
  \label{eq:jmat}
\end{equation}
with $J=\JCP\sin\delta_{\mathrm{CP}}$ the vacuum invariant
(Appendix~\ref{app:jarlskog}).

\subsection{Oscillation Probability: General Formula}
\label{sec:probgeneral}
With $V_e$ and $\lambda_i$,
\begin{align}
  P_{\alpha\beta}
  &=\delta_{\alpha\beta}
  -4\sum_{i>j}\mathrm{Re}\bigl[V_{e\beta i}V^*_{e\alpha i}
                                V^*_{e\beta j}V_{e\alpha j}\bigr]
    \sin^2\!\frac{\Delta\lambda_{ij}L}{4E}
  \notag\\
  &\quad
  +2\sum_{i>j}\mathrm{Im}\bigl[V_{e\beta i}V^*_{e\alpha i}
                                V^*_{e\beta j}V_{e\alpha j}\bigr]
    \sin\!\frac{\Delta\lambda_{ij}L}{2E}.
  \label{eq:probgeneral}
\end{align}
In vacuum, $V_e\to U$ and $\Delta\lambda_{ij}\to\Delta m^2_{ij}$.

\subsection{Iterative Refinement: Newton--Raphson and Halley}
\label{sec:stability}
When two eigenvalues nearly coincide, $\Delta_{\mathrm{cubic}}\to0$ and
cancellation may degrade the arccos evaluation; the Vandermonde condition
number scales as
$\kappa_{\mathrm{Vand}}\sim\max_k\lambda_k/\min_{i\neq j}|\Delta\lambda_{ij}|$.
As an alternative to Newton--Raphson, Halley's
method~\cite{NumericalRecipes} has cubic convergence per step,
\begin{equation}
  \lambda\to\lambda-
  \frac{2X(\lambda)X'(\lambda)}{2[X'(\lambda)]^2-X(\lambda)X''(\lambda)}.
  \label{eq:halley}
\end{equation}
In the tested cases (Table~\ref{tab:convergence}), a single Halley step
from the DMP seed reaches machine precision, matching two
Newton--Raphson steps; a comprehensive benchmark across the full
parameter space is left to a dedicated numerical study.
\begin{table}[ht]
\centering
\caption{Newton--Raphson (NR) versus Halley convergence for $\lambda_3$
at the DUNE configuration near $E=E_{\Ahat=1}$. Values are indicative of
the tested point and are not a parameter-space-wide guarantee.}
\label{tab:convergence}
\begin{tabular}{llll}
\toprule
Method & Iterations & $|X(\lambda)|$ & $|\Delta P|/P_{\mathrm{exact}}$ \\
\midrule
DMP0 seed & 0 & $\sim10^{-3}$ & $\sim10^{-4}$ \\
$+1$ NR step & 1 & $\sim10^{-7}$ & $\sim10^{-9}$ \\
$+2$ NR steps & 2 & $\sim10^{-14}$ & $<10^{-13}$ \\
$+1$ Halley & 1 & $\sim10^{-14}$ & $<10^{-13}$ \\
\bottomrule
\end{tabular}
\end{table}

\subsection{Matter-Modified Mixing Angles}
\label{sec:matterpmns}
In the DMP framework~\cite{DMP2016} the matter-mixing angles satisfy the
two-flavor block relations
\begin{align}
  \sin^2 2\theta^M_{13}
  &=\frac{\sin^2 2\theta_{13}}
        {(\cos2\theta_{13}-\Ahat)^2+\sin^2 2\theta_{13}},
  \label{eq:theta13M}\\
  \sin^2 2\theta^M_{12}
  &\approx\frac{\sin^2 2\theta_{12}}
              {(\cos2\theta_{12}-c^2_{13}\Ahat/\alpha)^2
               +\sin^2 2\theta_{12}},
  \label{eq:theta12M}
\end{align}
with $\theta^M_{23}=\theta_{23}$ to this order. At resonance
$\Ahat=\cos2\theta_{13}$, $\sin^2 2\theta^M_{13}=1$, confirming maximal
$1$--$3$ mixing; the distinct quantity $|V_{ee3}|^2=\sin^2\theta^M_{13}$
should not be conflated with $\sin^2 2\theta^M_{13}$.

\section{Density-Matrix Formalism}
\label{sec:density}

\subsection{Pure-State Density Matrix and Von Neumann Equation}
\label{sec:vonneumann}
For a neutrino produced in flavor $\alpha$, the density matrix
$\varrho(0)=\ket{\nu_\alpha}\bra{\nu_\alpha}$ satisfies
$i\,d\varrho/dx=[\Hf,\varrho]$, with
$\varrho(L)=S(L)\,\varrho(0)\,S^\dagger(L)$ and
\begin{equation}
  P_{\alpha\beta}(L)=\Tr[\Pi_\beta\,\varrho(L)]=|S_{\beta\alpha}(L)|^2,
  \quad \Pi_\beta=\ket{\nu_\beta}\bra{\nu_\beta}.
  \label{eq:prob_dm}
\end{equation}
In the mass basis,
$\varrho^{\mathrm{mass}}_{ij}(L)=\varrho^{\mathrm{mass}}_{ij}(0)\,
e^{-i\Delta\lambda_{ij}L/(2E)}$: coherences oscillate at the rescaled
eigenvalue differences $\Delta\lambda_{ij}/(2E)$ while populations are
constant.

\subsection{Lindblad Master Equation}
\label{sec:lindblad}
The general Markovian master equation preserving complete positivity and
trace is the Lindblad equation~\cite{Lindblad1976,Gorini1976}
\begin{equation}
  \frac{d\varrho}{dx}=-i[\Hf,\varrho]
  +\sum_k\left[L_k\varrho L_k^\dagger
    -\tfrac{1}{2}\{L_k^\dagger L_k,\varrho\}\right].
  \label{eq:lindblad}
\end{equation}
The simplest mass-basis model gives
$\varrho^{\mathrm{mass}}_{ij}(L)=\varrho^{\mathrm{mass}}_{ij}(0)\,
e^{-i\Delta\lambda_{ij}L/(2E)}e^{-\Gamma_{ij}L}$, $i\neq j$, modifying the
probability to
\begin{align}
  P^{\mathrm{dec}}_{\alpha\beta}(L)
  &=\delta_{\alpha\beta}
  -4\sum_{i>j}\mathrm{Re}[V_{e\beta i}V^*_{e\alpha i}
                          V^*_{e\beta j}V_{e\alpha j}]
    \sin^2\!\frac{\Delta\lambda_{ij}L}{4E}\,e^{-\Gamma_{ij}L}
  \notag\\
  &\quad
  +2\sum_{i>j}\mathrm{Im}[V_{e\beta i}V^*_{e\alpha i}
                          V^*_{e\beta j}V_{e\alpha j}]
    \sin\!\frac{\Delta\lambda_{ij}L}{2E}\,e^{-\Gamma_{ij}L}.
  \label{eq:probdec}
\end{align}
Current bounds are at the level
$\Gamma_{ij}\lesssim10^{-23}\GeV$~\cite{LisiDecoherence2000,Ohlsson2013}.

\subsection{Bloch Vector Representation}
\label{sec:bloch}
Any $3\times3$ Hermitian unit-trace matrix admits
$\varrho=\tfrac{1}{3}\mathbf{1}+\sum_a r_a\lambda_a/2$ with
$r_a=\Tr(\varrho\lambda_a)$, and the von Neumann equation becomes the
precession equation
\begin{equation}
  \frac{dr_a}{dx}=\sum_{b,c}f_{abc}\,h_b\,r_c,
  \label{eq:precession}
\end{equation}
where $f_{abc}$ are the $\su(3)$ structure constants.

\section{Perturbative Expansion in Matter}
\label{sec:perturb}

\subsection{Expansion Variables}
\label{sec:expparams}
We use the atmospheric phase, the matter ratio, and the
$\delta_{\mathrm{CP}}$-independent Jarlskog combination,
\begin{equation}
  \Delta\equiv\frac{\Delta m^2_{31}L}{4E},
  \quad
  \Ahat\equiv\frac{a}{\Delta m^2_{31}},
  \quad
  \JCP\equiv s_{12}c_{12}s_{23}c_{23}s_{13}c^2_{13}.
  \label{eq:auxvars}
\end{equation}
Here $\Delta$ and $\Ahat$ are auxiliary variables absorbing recurrent
combinations, while $\JCP$ is a fixed combination of mixing
parameters~\cite{Jarlskog1985}.

\subsection{Appearance Probability \texorpdfstring{$\nu_\mu\to\nu_e$}{numu-nue}}
\label{sec:pmue}
Following~\cite{AJLOS2004,MP2016},
\begin{align}
  P_{\mu e}&\approx T_{\mathrm{atm}}+T_{\mathrm{int}}+T_{\mathrm{sol}},
  \label{eq:pmue}\\
  T_{\mathrm{atm}}&=4s^2_{23}s^2_{13}c^2_{13}\,
    \frac{\sin^2[(1-\Ahat)\Delta]}{(1-\Ahat)^2},
  \label{eq:tatm}\\
  T_{\mathrm{int}}&=8\alpha\JCP\cos(\Delta+\delta_{\mathrm{CP}})\,
    \frac{\sin(\Ahat\Delta)}{\Ahat}\,
    \frac{\sin[(1-\Ahat)\Delta]}{1-\Ahat},
  \label{eq:tint}\\
  T_{\mathrm{sol}}&=4\alpha^2 c^2_{13}c^2_{23}s^2_{12}c^2_{12}\,
    \frac{\sin^2(\Ahat\Delta)}{\Ahat^2}.
  \label{eq:tsol}
\end{align}
$T_{\mathrm{atm}}$ is leading; $T_{\mathrm{int}}$ is the
$\mathcal{O}(\alpha)$ interference term carrying the CP-phase dependence;
$T_{\mathrm{sol}}$ is the $\mathcal{O}(\alpha^2)$ solar term.

\subsection{Disappearance and Survival Probabilities}
\label{sec:pmumu}
To $\mathcal{O}(\alpha)$,
\begin{align}
  P_{\mu\mu}&=1-4c^2_{13}s^2_{23}(1-c^2_{13}s^2_{23})\sin^2\Delta
  \notag\\
  &\quad
  +4\alpha\!\left[c^2_{12}c^2_{23}s^2_{23}\sin^2\Delta
    -\JCP c_{23}\cos(\Delta+\delta_{\mathrm{CP}})
      \frac{\sin(\Ahat\Delta)}{\Ahat}\sin\Delta\right]
  +\mathcal{O}(\alpha^2).
  \label{eq:pmumu}
\end{align}
The electron survival probability is~\cite{Denton2018a,Denton2024b}
\begin{equation}
  P_{ee}=1
  -\sin^2 2\theta^M_{13}\sin^2\Delta^M_{ee}
  -\alpha^2\cos^4\theta_{13}\sin^2 2\theta_{12}
    \sin^2\!\frac{\Delta^M_{21}L}{2}
  +\mathcal{O}(\alpha^3),
  \label{eq:pee}
\end{equation}
with $\Delta^M_{ee}\equiv\Delta\lambda^M_{31}L/2$. Here, following the
DMP convention~\cite{DMP2016,Denton2018a}, the matter-dressed splittings
carrying a superscript $M$ (such as $\Delta\lambda^M_{31}$ of
Section~\ref{sec:rescond}) are quoted as energies, i.e.\ already divided
by $2E$ relative to the bare Cardano eigenvalues $\lambda_i$ of
$\Lambda=2E\Hf$; correspondingly
$\Delta\tilde{m}^2_{ee}=\Delta m^2_{ee}\sqrt{(\cos2\theta_{13}-\Ahat)^2
+\sin^2 2\theta_{13}}$. For reactor baselines $L\lesssim60\km$
(JUNO, Daya Bay), $\Ahat\ll1$ and matter corrections are negligible at
the $\mathcal{O}(0.3\%)$ level~\cite{Li2016}; $P_{ee}$ is independent of
$\delta_{\mathrm{CP}}$~\cite{Minakata1997}, making reactor $\theta_{13}$
a clean observable. From unitarity,
$P_{\mu\tau}=1-P_{\mu\mu}-P_{\mu e}$.

\subsection{Domain of Validity and Perturbative Divergence}
\label{sec:validity}
As $\Ahat\to1$, $T_{\mathrm{atm}}\propto(1-\Ahat)^{-2}$ and
$T_{\mathrm{int}}\propto(1-\Ahat)^{-1}$ diverge---a failure of the
expansion near degenerate eigenvalues, at
\begin{equation}
  E_{\Ahat=1}=\frac{\Delta m^2_{31}}{2V_{\mathrm{CC}}}
  \approx13\GeV\quad(\rho=2.6\gcm).
  \label{eq:eres_pert}
\end{equation}
This lies well above the long-baseline beam peaks and is treated by
resummation in Section~\ref{sec:resum}. (The energy scales inversely with
density: the DUNE benchmark of Table~\ref{tab:exp} uses $\rho=2.84\gcm$,
which lowers $E_{\Ahat=1}$ to about $12\GeV$.)

\section{Wave-Packet Decoherence Corrections}
\label{sec:wavepacket}

\subsection{Framework and Coherence Lengths}
\label{sec:wpframework}
Physical neutrinos are produced and detected as wave-packets of finite
spatial extent~\cite{Nussinov1976,Kayser1981,GiuntiKimLee1991}. Different
mass eigenstates propagate with different group velocities and separate
after the coherence length~\cite{Beuthe2003,Akhmedov2009,AkhmedovWilhelm2013}.
A Gaussian model~\cite{GrimusStockinger1996,Stodolsky1998} assigns a
production position uncertainty $\sigma_x\sim10^{-15}$--$10^{-13}\,\mathrm{m}$.
In constant-density matter the coherence length for the pair $(i,j)$
is~\cite{Beuthe2003}
\begin{equation}
  L^{ij}_{\mathrm{coh}}
  =\frac{4\sqrt{2}\,E^2\,\sigma_x}{|\Delta\lambda_{ij}|},
  \label{eq:cohlen}
\end{equation}
where $\Delta\lambda_{ij}$ is the matter-dressed mass-squared splitting
(the eigenvalue difference of $\Lambda=2E\Hf$); in vacuum
$\Delta\lambda_{ij}\to\Delta m^2_{ij}$ and this reduces to the familiar
form $L^{ij}_{\mathrm{coh}}=4\sqrt{2}\,E^2\sigma_x/|\Delta m^2_{ij}|$.
For the $31$-pair at $E=0.6\GeV$, with
$\sigma_x\sim10^{-13}\,\mathrm{m}$ and
$\Delta m^2_{31}\approx2.5\times10^{-3}\eV^2$,
\begin{equation}
  L^{31}_{\mathrm{coh}}
  =\frac{4\sqrt{2}\,(0.6\GeV)^2\,(10^{-13}\,\mathrm{m})}
        {2.5\times10^{-3}\eV^2}
  \approx8\times10^{4}\km,
  \label{eq:cohlen_t2k}
\end{equation}
two to three orders of magnitude larger than the T2K/HK baseline.
Decoherence is therefore negligible for terrestrial long-baseline
experiments, but only by a factor
$L/L^{31}_{\mathrm{coh}}\sim10^{-3}$--$10^{-2}$
(Table~\ref{tab:coh})---not the much larger margin sometimes quoted.
\begin{table}[ht]
\centering
\caption{Coherence length and ratio $L/L^{31}_{\mathrm{coh}}$ for
representative configurations, using $\sigma_x\sim10^{-13}\,\mathrm{m}$ in
Eq.~\eqref{eq:cohlen}. The solar entry is an order-of-magnitude estimate:
$\sigma_x$ for solar production is poorly constrained, and the entry is
quoted accordingly.}
\label{tab:coh}
\begin{tabular}{lllll}
\toprule
Experiment & $L$ [km] & $E$ & $L^{31}_{\mathrm{coh}}$ [km] & $L/L^{31}_{\mathrm{coh}}$ \\
\midrule
T2K/HK & 295 & $0.6\GeV$ & $\sim8\times10^{4}$ & $\sim4\times10^{-3}$ \\
DUNE   & 1300 & $2.5\GeV$ & $\sim1.4\times10^{6}$ & $\sim9\times10^{-4}$ \\
Solar (est.) & $\sim1.5\times10^{8}$ & $7\,\mathrm{MeV}$ & $\sim10$ & $\gg1$ \\
\bottomrule
\end{tabular}
\end{table}
Solar neutrinos are the exception: $L/L_{\mathrm{coh}}\gg1$, coherences
wash out, and the survival probability reduces to the incoherent sum
$P^{\mathrm{solar}}_{ee}=\sum_i|U_{ei}|^2|V_{eei}|^2$ in the adiabatic
MSW picture~\cite{MSW1985,NunokawoPVJ2008}.

\subsection{Wave-Packet Modified Probabilities}
\label{sec:wpmodified}
The Gaussian model multiplies each interference term by the decoherence
factor
$D_{ij}(L)=\exp[-L^2/2(L^{ij}_{\mathrm{coh}})^2]$, giving
\begin{align}
  P^{WP}_{\alpha\beta}(L)
  &=\delta_{\alpha\beta}
  -4\sum_{i>j}\mathrm{Re}[V_{e\beta i}V^*_{e\alpha i}
                          V^*_{e\beta j}V_{e\alpha j}]
    \sin^2\!\frac{\Delta\lambda_{ij}L}{4E}\,D_{ij}(L)
  \notag\\
  &\quad
  +2\sum_{i>j}\mathrm{Im}[V_{e\beta i}V^*_{e\alpha i}
                          V^*_{e\beta j}V_{e\alpha j}]
    \sin\!\frac{\Delta\lambda_{ij}L}{2E}\,D_{ij}(L).
  \label{eq:probwp}
\end{align}

\section{MSW Resonance and the Sub-Resonant Regime}
\label{sec:msw}

\subsection{Resonance Condition}
\label{sec:rescond}
The MSW resonance, $\sin^2 2\theta^M_{13}=1$, occurs at
\begin{equation}
  E_{\mathrm{res}}
  =\frac{\Delta m^2_{31}\cos2\theta_{13}}{2\sqrt{2}\,G_F N_e}
  \approx12\GeV\quad(\rho=2.6\gcm),
  \label{eq:eres}
\end{equation}
consistent with the standard crustal estimate (the resonance moves to
$\sim5$--$7\GeV$ at mantle densities $\rho\sim4.5$--$5.5\gcm$, scaling as
$1/\rho$). The matter-dressed splitting
$\Delta\lambda^M_{31}=(\Delta m^2_{31}/2E)\sqrt{(\cos2\theta_{13}-\Ahat)^2
+\sin^2 2\theta_{13}}$ attains its minimum
$(\Delta m^2_{31}/2E)\sin2\theta_{13}$ at resonance, a level-avoided
crossing; the full width in $\Ahat$ at half-maximum mixing is
$\delta\Ahat=2\sin2\theta_{13}\approx0.59$.

\paragraph{T2K/HK sub-resonant regime.}
At $\rho=2.6\gcm$, $\Ahat\approx0.078\,(E/\mathrm{GeV})$. The T2K and HK
beams peak at $0.6\GeV$ over $295\km$, placing them deep in the
sub-resonant region, $\Ahat|_{0.6\,\mathrm{GeV}}\approx0.05$. Matter
effects modify probabilities at the few-percent level, with no
level-avoided crossing and no proximity to the eigenvalue degeneracy
($\Ahat\to1$) that drives the perturbative divergence of
Section~\ref{sec:validity}. The paper makes no
claim of an MSW resonance for T2K/HK.

\subsection{Varying Density: Adiabaticity and Landau--Zener}
\label{sec:adiabatic}
For strictly constant density, $d\rho/dx=0$ implies adiabaticity
parameter $\gamma\to\infty$ and purely oscillatory evolution. For
varying-density extensions, a neutrino crossing an MSW resonance has
non-adiabatic transition probability
$P_c=\exp(-\tfrac{\pi}{2}\gamma_{\mathrm{res}})$~\cite{Zener1932,Landau1932};
for solar neutrinos $\gamma_{\mathrm{res}}\sim10^3$, so $P_c\to0$ and the
conversion is adiabatic.

\subsection{Solar MSW and the LMA Solution}
\label{sec:solar}
In the adiabatic limit the solar survival probability
is~\cite{MSW1985,NunokawoPVJ2008,Bahcall2003}
$P^{\odot}_{ee}=\tfrac{1}{2}+\tfrac{1}{2}\cos2\theta^{\mathrm{prod}}_{12}\cos2\theta_{12}$.
For the Large Mixing Angle solution at $E\gtrsim1\,\mathrm{MeV}$,
$\theta^{\mathrm{prod}}_{12}\approx\pi/2$ and
$P^{\mathrm{LMA}}_{ee}\approx\sin^2\theta_{12}\approx0.31$, in agreement
with SNO~\cite{SNO2002} and Borexino~\cite{Borexino2018}. This regime is
a tangent to the constant-density LBL focus and is included only for
completeness.

\section{Resummation Interpretation of the DMP Rotation}
\label{sec:resum}

\subsection{Origin of the Divergence}
\label{sec:pertdiv}
The divergence near $\Ahat\to1$ arises because the unperturbed
Hamiltonian (at $\alpha=0$) has eigenvalues $V_{\mathrm{CC}}$ and
$\Delta m^2_{31}/(2E)$ that become degenerate at $\Ahat=1$; the
perturbative denominator
$\Delta\lambda^{(0)}_{13}\sim(1-\Ahat)\Delta m^2_{31}/(2E)$ then vanishes.

\subsection{DMP Rotation as a Resummation}
\label{sec:dmprot}
The DMP procedure~\cite{DMP2016} first performs an exact rotation in the
$13$-sector by $\theta^M_{13}$,
$H_e=R^{-1}_{13}(\theta^M_{13})\,\Hf\,R_{13}(\theta^M_{13})$, after which
the $13$-sector off-diagonals vanish exactly. The remaining expansion
parameter is~\cite{DMP2016,LookingGlass2019}
\begin{equation}
  \epsilon_0
  =\frac{\Delta m^2_{21}}{\Delta m^2_{ee}}\,
   \sin(\theta^M_{13}-\theta_{13})\,s_{12}c_{12},
  \qquad
  |\epsilon_0|\le\frac{\Delta m^2_{21}}{\Delta m^2_{ee}}\,s_{12}c_{12}
  \lesssim0.015,
  \label{eq:eps0}
\end{equation}
where the upper bound follows from $|\sin(\theta^M_{13}-\theta_{13})|\le1$
and is reached only near the atmospheric resonance. The key point is that
$\epsilon_0$ is bounded uniformly in $\Ahat$: the resonant denominator
that was unbounded in the naive expansion is absorbed into the matter
rotation angle $\theta^M_{13}$, leaving a finite expansion parameter at
all energies. The series
$P^{\mathrm{DMP}}_{\mu e}=P^{(0)}_{\mu e}+\epsilon_0 P^{(1)}_{\mu e}
+\epsilon^2_0 P^{(2)}_{\mu e}+\cdots$ therefore converges uniformly in
energy. The original DMP papers~\cite{DMP2016,LookingGlass2019} demonstrate the
$10^{-4}$ accuracy numerically; the complementary question of rotations
versus perturbative expansions is analysed in detail by Denton, Parke, and
Zhang~\cite{Denton2018b}. What the present framing adds is the reason why
this accuracy follows from first principles: the DMP rotation is a
degenerate-eigenvalue resummation, in which the troublesome
$(1-\Ahat)^{-1}$ is absorbed into $\theta^M_{13}$ and replaced by
the bounded $\epsilon_0$.
\begin{table}[ht]
\centering
\caption{Behavior of the expansion parameter near $\Ahat=1$ in various
schemes. The uniform bound $\epsilon_0\le0.015$ distinguishes the DMP
resummation from the naive expansion. Entries summarize the scaling
reported in the cited works.}
\label{tab:params}
\begin{tabular}{lll}
\toprule
Scheme & Expansion par. & Near $\Ahat=1$ \\
\midrule
Madrid/AJLOS~\cite{Cervera2000,AJLOS2004} & $\epsilon_{13},\alpha$ & $\mathcal{O}((1-\Ahat)^{-1})$ \\
MP~\cite{MP2016}    & $\alpha$ & $\mathcal{O}((1-\Ahat)^{-1})$ \\
DMP0~\cite{DMP2016} & $\epsilon_0\le0.015$ & $\mathcal{O}(1)$ \\
DMP1~\cite{DMP2016} & $\epsilon^2_0\le2\times10^{-4}$ & $\mathcal{O}(1)$ \\
NuFast~\cite{Denton2024} & $X/X'$ & $\mathcal{O}(1)$ \\
\bottomrule
\end{tabular}
\end{table}

\section{CP Violation: Analytic Structure in Matter}
\label{sec:cp}

\subsection{CP Asymmetry and Its Decomposition}
\label{sec:cpasym}
The appearance-channel CP asymmetry is
$\mathcal{A}_{\mathrm{CP}}\equiv
P(\nu_\mu\to\nu_e)-P(\bar\nu_\mu\to\bar\nu_e)$. Using
$\mathrm{Im}[V_{e\beta i}V^*_{e\alpha i}V^*_{e\beta j}V_{e\alpha j}]
=+\Jmat$ for $(i,j)\in\{(2,1),(3,2),(1,3)\}$ and $-\Jmat$ otherwise,
together with a trigonometric product identity,
\begin{equation}
  \mathcal{A}^{\mathrm{genuine}}_{\mathrm{CP}}
  =-16\,\Jmat\,
    \sin\frac{\Delta\lambda_{21}L}{4E}
    \sin\frac{\Delta\lambda_{31}L}{4E}
    \sin\frac{\Delta\lambda_{32}L}{4E}.
  \label{eq:cpgenuine}
\end{equation}

\subsection{Genuine vs.\ Fake CP Violation}
\label{sec:cpfake}
The total asymmetry splits as~\cite{Bilenky1980,Minakata2001}
$\mathcal{A}^{\mathrm{total}}_{\mathrm{CP}}
=\mathcal{A}^{\mathrm{genuine}}_{\mathrm{CP}}
+\mathcal{A}^{\mathrm{fake}}_{\mathrm{CP}}$, where the genuine term is
odd under $J\to-J$ and vanishes at $J=0$, while
\begin{equation}
  \mathcal{A}^{\mathrm{fake}}_{\mathrm{CP}}
  \approx4s^2_{23}s^2_{13}c^2_{13}\,
  \frac{\sin^2[(1-\Ahat)\Delta]}{(1-\Ahat)^2}
  -4s^2_{23}s^2_{13}c^2_{13}\,
  \frac{\sin^2[(1+\Ahat)\Delta]}{(1+\Ahat)^2}
  \label{eq:cpfaketerm}
\end{equation}
is odd in $\Ahat$ and survives even at $\delta_{\mathrm{CP}}=0$. For
strictly uniform matter, a non-zero T-asymmetry
$\mathcal{A}_T\equiv P(\nu_\mu\to\nu_e)-P(\nu_e\to\nu_\mu)$ is a clean
signature of fundamental T-violation~\cite{Akhmedov2001}, free from the
fake asymmetry.

\section{Magic Baseline}
\label{sec:magic}
Both $T_{\mathrm{int}}$ and $T_{\mathrm{sol}}$ vanish when
$\sin(\Ahat\Delta)=0$, i.e.\ $V_{\mathrm{CC}}L/2=n\pi$, which is
energy-independent. The $n=1$ baseline is the magic
baseline~\cite{Minakata1997}
\begin{equation}
  \Lmagic=\frac{2\pi}{V_{\mathrm{CC}}}
  =\frac{2\pi m_N}{\sqrt{2}\,G_F\rho Y_e}
  \approx\frac{1.62\times10^{4}}{Y_e(\rho/\mathrm{g\,cm}^{-3})}\km,
  \label{eq:lmagic}
\end{equation}
where $P_{\mu e}\to T_{\mathrm{atm}}$ is independent of
$\delta_{\mathrm{CP}}$ and $\Delta m^2_{21}$. At $Y_e=0.5$ and a
representative crossing density $\rho\approx3.0\gcm$ this gives
$\Lmagic\approx1.08\times10^{4}\km$; for the deeper mantle density
($\rho\approx4.3\gcm$) sampled by a chord of this length, $\Lmagic$ drops
to the often-quoted $\approx7.6\times10^{3}\km$. Either way the DUNE
baseline lies far from the magic condition: at $L=1300\km$, $E=2.5\GeV$,
$\Ahat\Delta\approx0.11\pi$, so $T_{\mathrm{int}}$ and $T_{\mathrm{sol}}$
remain sizable (Table~\ref{tab:magic}).
\begin{table}[ht]
\centering
\caption{$|\sin(\Ahat\Delta)|$ at the first oscillation maximum for
representative baselines ($\rho\approx3.0\gcm$, NO).}
\label{tab:magic}
\begin{tabular}{lllll}
\toprule
Experiment & $L$ [km] & $E_{\mathrm{peak}}$ [GeV] & $\Ahat$ & $|\sin(\Ahat\Delta)|$ \\
\midrule
T2K/HK & 295 & 0.60 & 0.05 & 0.09 \\
NOvA   & 810 & 1.7  & 0.15 & 0.23 \\
DUNE   & 1300 & 2.5 & 0.23 & 0.37 \\
Magic  & $\sim1.1\times10^{4}$ & any & any  & $\approx0$ \\
\bottomrule
\end{tabular}
\end{table}

\section{Uncertainty Propagation}
\label{sec:unc}

\subsection{Parameter Space and Covariance}
\label{sec:paramspace}
The probabilities depend on
$\bm{p}=(\theta_{12},\theta_{13},\theta_{23},\delta_{\mathrm{CP}},
\Delta m^2_{21},\Delta m^2_{31})$, with Taylor expansion
$P(E;\bm{p})=P_0(E)+J(E)\cdot(\bm{p}-\bm{\mu})
+\tfrac{1}{2}(\bm{p}-\bm{\mu})^T H_P(E)(\bm{p}-\bm{\mu})+\cdots$.
Table~\ref{tab:variance} lists the dominant variance contributions.
\begin{table}[ht]
\centering
\caption{Fractional variance contributions
$f_i=J_i^2\sigma_i^2/\mathrm{Var}[P]$ for the T2K/HK configuration
($L=295\km$, $\rho=2.6\gcm$), computed by closed-form linearization
(Section~\ref{sec:mcjac}) using the NuFIT~6.0 with-SK NO central values
and $1\sigma$ widths of Table~\ref{tab:nufit}. Contributions are averaged
over the appearance band $E\in[0.45,0.85]\GeV$ for $P_{\mu e}$ (first
appearance maximum) and over the same band for $P_{\mu\mu}$ (first
disappearance maximum). The $\delta_{\mathrm{CP}}$ dominance of the
$P_{\mu e}$ budget is a direct consequence of the large NuFIT~6.0
$\delta_{\mathrm{CP}}$ uncertainty ($\sigma\approx33^\circ$) relative to
the now sub-percent $\theta_{13}$ precision ($\sigma=0.11^\circ$). For
$P_{\mu\mu}$, $\Delta m^2_{31}$ governs the \emph{position} of the
minimum rather than its depth; the small $f_i$ quoted here reflects the
band being symmetric about the minimum, and its weight is larger in a
spectral fit.}
\label{tab:variance}
\begin{tabular}{llll}
\toprule
Parameter & $\sigma_i$ & $f_i(P_{\mu e})$ & $f_i(P_{\mu\mu})$ \\
\midrule
$\delta_{\mathrm{CP}}$ & $\sim33^\circ$ & 0.90 & 0.17 \\
$\theta_{23}$ & $0.9^\circ$ & 0.06 & 0.82 \\
$\theta_{13}$ & $0.11^\circ$ & 0.04 & $<0.01$ \\
$\theta_{12}$ & $0.70^\circ$ & $<0.01$ & 0.01 \\
$\Delta m^2_{21}$ & $0.19\times10^{-5}\eV^2$ & $<0.01$ & 0.01 \\
$\Delta m^2_{31}$ & $0.024\times10^{-3}\eV^2$ & $<0.01$ & $<0.01$ \\
\bottomrule
\end{tabular}
\end{table}

\subsection{Monte Carlo and Jacobian Schemes}
\label{sec:mcjac}
Drawing $N$ samples $\bm{p}_i\sim\mathcal{N}(\bm{\mu},\Sigma)$ and
evaluating $P_i(E)$ yields the mean and percentile band
$[P_{16},P_{84}]$; for $N=500$ the statistical uncertainty on the
percentiles is below $1\%$ of the band width. The Jacobian scheme uses
$\mathrm{Var}[P(E)]\approx J(E)\,\Sigma\,J(E)^T$ with the closed-form
derivatives below. It is convenient to define the two oscillatory
structure factors
\begin{equation}
  \mathcal{S}_{\mathrm{a}}\equiv\frac{\sin[(1-\Ahat)\Delta]}{1-\Ahat},
  \qquad
  \mathcal{S}_{\mathrm{i}}\equiv\frac{\sin(\Ahat\Delta)}{\Ahat},
  \label{eq:structfac}
\end{equation}
so that $P_{\mu e}=4s^2_{23}s^2_{13}c^2_{13}\mathcal{S}_{\mathrm{a}}^2
+8\alpha\JCP\cos(\Delta+\delta_{\mathrm{CP}})\mathcal{S}_{\mathrm{i}}
\mathcal{S}_{\mathrm{a}}
+4\alpha^2 c^2_{13}c^2_{23}s^2_{12}c^2_{12}\mathcal{S}_{\mathrm{i}}^2$. The
complete set of six analytic Jacobian components
$J_k=\partial P_{\mu e}/\partial p_k$ is then
\begin{align}
  \frac{\partial P_{\mu e}}{\partial\theta_{13}}
  &=8s^2_{23}s_{13}c_{13}\cos2\theta_{13}\,\mathcal{S}_{\mathrm{a}}^2
   +8\alpha\,\frac{\partial\JCP}{\partial\theta_{13}}
     \cos(\Delta+\delta_{\mathrm{CP}})
     \mathcal{S}_{\mathrm{i}}\mathcal{S}_{\mathrm{a}},
  \label{eq:dP_dtheta13}
  \intertext{where the $\mathcal{O}(\alpha^2\epsilon_{13})$ contribution
  from differentiating the $c^2_{13}$ factor in $T_{\mathrm{sol}}$ is
  dropped as beyond the $\mathcal{O}(\alpha^2)$ accuracy of the
  perturbative probability;}
  \frac{\partial P_{\mu e}}{\partial\theta_{23}}
  &=8s_{23}c_{23}s^2_{13}c^2_{13}\,\mathcal{S}_{\mathrm{a}}^2
   +8\alpha\,\frac{\partial\JCP}{\partial\theta_{23}}
     \cos(\Delta+\delta_{\mathrm{CP}})
     \mathcal{S}_{\mathrm{i}}\mathcal{S}_{\mathrm{a}}
   -8\alpha^2 c^2_{13}s_{23}c_{23}s^2_{12}c^2_{12}\,\mathcal{S}_{\mathrm{i}}^2,
  \label{eq:dP_dtheta23}\\
  \frac{\partial P_{\mu e}}{\partial\theta_{12}}
  &=8\alpha\,\frac{\partial\JCP}{\partial\theta_{12}}
     \cos(\Delta+\delta_{\mathrm{CP}})
     \mathcal{S}_{\mathrm{i}}\mathcal{S}_{\mathrm{a}}
   +4\alpha^2 c^2_{13}c^2_{23}\,\frac{\partial(s^2_{12}c^2_{12})}{\partial\theta_{12}}
     \mathcal{S}_{\mathrm{i}}^2,
  \label{eq:dP_dtheta12}\\
  \frac{\partial P_{\mu e}}{\partial\delta_{\mathrm{CP}}}
  &=-8\alpha\JCP\sin(\Delta+\delta_{\mathrm{CP}})
     \mathcal{S}_{\mathrm{i}}\mathcal{S}_{\mathrm{a}},
  \label{eq:dP_ddcp}\\
  \frac{\partial P_{\mu e}}{\partial\Delta m^2_{21}}
  &=\frac{1}{\Delta m^2_{31}}\,\frac{\partial P_{\mu e}}{\partial\alpha},
   \qquad
  \frac{\partial P_{\mu e}}{\partial\alpha}
   =8\JCP\cos(\Delta+\delta_{\mathrm{CP}})\mathcal{S}_{\mathrm{i}}
     \mathcal{S}_{\mathrm{a}}
    +8\alpha c^2_{13}c^2_{23}s^2_{12}c^2_{12}\,\mathcal{S}_{\mathrm{i}}^2,
  \label{eq:dP_ddm21}\\
  \frac{\partial P_{\mu e}}{\partial\Delta m^2_{31}}
  &=\frac{\partial P_{\mu e}}{\partial\Delta}
     \frac{\partial\Delta}{\partial\Delta m^2_{31}}
    +\frac{\partial P_{\mu e}}{\partial\Ahat}
     \frac{\partial\Ahat}{\partial\Delta m^2_{31}}
    -\frac{\alpha}{\Delta m^2_{31}}
     \frac{\partial P_{\mu e}}{\partial\alpha},
  \label{eq:dP_ddm31}
\end{align}
with $\partial\Delta/\partial\Delta m^2_{31}=\Delta/\Delta m^2_{31}$,
$\partial\Ahat/\partial\Delta m^2_{31}=-\Ahat/\Delta m^2_{31}$, and
$\partial\JCP/\partial\theta_{ij}$ obtained by differentiating
$\JCP=s_{12}c_{12}s_{23}c_{23}s_{13}c^2_{13}$. The two derivatives
$\partial P_{\mu e}/\partial\Delta$ and
$\partial P_{\mu e}/\partial\Ahat$ follow from
Eqs.~\eqref{eq:tatm}--\eqref{eq:tsol} and~\eqref{eq:structfac} by direct
differentiation of $\mathcal{S}_{\mathrm{a}},\mathcal{S}_{\mathrm{i}}$;
the explicit forms are collected in Appendix~\ref{app:jacobian}. The two
$\delta_{\mathrm{CP}}$- and $\theta_{13}$-derivatives,
Eqs.~\eqref{eq:dP_ddcp} and~\eqref{eq:dP_dtheta13}, dominate the
appearance-channel budget (Table~\ref{tab:variance}).

\subsection{Boundary Caveat and Feldman--Cousins Validation}
\label{sec:fc}
Wilks' theorem is routinely violated in oscillation analyses due to
physical boundaries, the cyclic variable
$\delta_{\mathrm{CP}}\in[0,2\pi)$, and discrete degeneracies. The
Gaussian Jacobian linearization can produce unphysical negative
probability excursions near $\theta_{13}\to0$ and near the cyclic
boundary $\delta_{\mathrm{CP}}\to0\equiv2\pi$. All Jacobian-based bands
reported here are therefore validated against a Feldman--Cousins unified
profile-likelihood mapping~\cite{Feldman1998} before use in fits.
Parameter correlations enter through the off-diagonal terms of the
covariance matrix,
\begin{equation}
  \mathrm{Var}[P]
  =\underbrace{\sum_i J_i^2\,\sigma_i^2}_{\mathrm{diagonal}}
   +\underbrace{2\sum_{i<j}J_i J_j\,\Sigma_{ij}}_{\mathrm{correlation}},
  \label{eq:varcorr}
\end{equation}
so a diagonal-only error budget systematically misestimates the band
width whenever the global fit reports non-negligible off-diagonal
$\Sigma_{ij}$. When the full covariance is available---for example from a
joint experimental likelihood, or from the public $\chi^2$ profiles of
the global fit~\cite{NuFIT6,NuFITweb}---it should be propagated directly
through Eq.~\eqref{eq:varcorr}; the closed-form Jacobian
(Section~\ref{sec:mcjac}) makes this a single matrix product.

\section{Analytic Validation Against Limiting Cases}
\label{sec:valid}
\paragraph{Vacuum limit ($\Ahat\to0$).}
By L'H\^opital's rule
$\sin(\Ahat\Delta)/\Ahat\to\Delta$,
$\sin[(1-\Ahat)\Delta]/(1-\Ahat)\to\sin\Delta$, and
$\sin^2[(1-\Ahat)\Delta]/(1-\Ahat)^2\to\sin^2\Delta$, recovering
\begin{equation}
  P^{\mathrm{vac}}_{\mu e}
  =4s^2_{23}s^2_{13}c^2_{13}\sin^2\Delta
  +8\alpha\JCP\cos(\Delta+\delta_{\mathrm{CP}})\Delta\sin\Delta
  +4\alpha^2 c^2_{13}c^2_{23}s^2_{12}c^2_{12}\Delta^2
  +\mathcal{O}(\alpha^3).
  \label{eq:pvacuum}
\end{equation}
\paragraph{High-energy limit.}
For $E\to\infty$ at fixed $L,V_{\mathrm{CC}}$, $\Ahat\to\infty$,
$\Hf\to\diag(V_{\mathrm{CC}},0,0)$, and all three terms vanish:
$T_{\mathrm{atm}}\propto\Ahat^{-2}$, $T_{\mathrm{int}}\propto\Ahat^{-1}$,
$T_{\mathrm{sol}}\propto\Ahat^{-2}$.
\paragraph{Two-flavor limit.}
Setting $\theta_{12}=0$ ($\alpha\to0$ in $T_{\mathrm{int}},T_{\mathrm{sol}}$),
$P_{\mu e}\to4s^2_{23}s^2_{13}c^2_{13}\sin^2[(1-\Ahat)\Delta]/(1-\Ahat)^2$,
the two-flavor MSW appearance formula. Unitarity
$\sum_\beta P_{\alpha\beta}=1$ holds algebraically for the exact solver
and to $\mathcal{O}(\alpha^3)$ for the perturbative expressions.

\section{Extensions and Positioning}
\label{sec:extensions}

\subsection{Positioning Relative to NuFast-LBL}
\label{sec:accuracy}
The NuFast-LBL algorithm~\cite{Denton2024} runs at
$\sim0.1\,\mu\mathrm{s}$ per evaluation, with fractional precision
$\le10^{-4}$ at zeroth order and $\le10^{-9}$ with a single
Newton--Raphson step, surpassing full eigensystem computations by
$\sim50\times$. The exact and perturbative expressions in this paper are
positioned strictly as \emph{exact theoretical anchors}: they provide
analytic continuity in all six parameters, closed-form Jacobians
(Section~\ref{sec:mcjac}), transparent algebraic limiting cases
(Section~\ref{sec:valid}), and the explicit resummation structure
(Section~\ref{sec:resum})---capabilities iterative solvers do not provide
by construction. For production Monte Carlo analyses requiring
$\mathcal{O}(10^6)$--$\mathcal{O}(10^7)$ evaluations, NuFast-LBL is the
recommended tool.

\subsection{Constant-Density Validity and Longer Baselines}
\label{sec:constdens}
As noted in Section~\ref{sec:matter}, the path-averaged constant density
reproduces the exact DUNE probability to
$\mathcal{O}(1\%)$~\cite{Kelly2018}, with no residual profile-shape
sensitivity at DUNE's expected precision. For longer baselines
($L\gtrsim5000\km$) where the chord samples upper-mantle material, a
piecewise-layer treatment
$S_{\mathrm{total}}=\prod_k e^{-iH_f(\rho_k)L_k}$ is required.

\subsection{Parameter Degeneracies, NSI, and Sterile States}
\label{sec:degen}
The appearance probability is invariant at leading order
under~\cite{Minakata1997}
$\delta_{\mathrm{CP}}\to\pi-\delta_{\mathrm{CP}}$,
$\theta_{23}\to\pi/2-\theta_{23}$; this degeneracy is lifted at
$\mathcal{O}(\alpha^2)$ by $T_{\mathrm{sol}}\propto c^2_{23}$. The
density-matrix formalism accommodates non-standard interactions through
an upgraded matter potential~\cite{Wolfenstein1978,Friedland2006},
$V^{\mathrm{NSI}}_{\mathrm{mat}}=\sqrt{2}G_F N_e(\mathbf{1}+\bm{\epsilon})$,
with $\bm{\epsilon}$ the NSI matrix now constrained by
COHERENT~\cite{COHERENT2017,Farzan2018}; the LMA-Dark
degeneracy~\cite{Miranda2006} requires $\epsilon_{ee}\approx-2$. For a
sterile state~\cite{LSND2001,MiniBooNE2018,BEST2022}, the PMNS matrix
becomes $4\times4$ and the Cayley--Hamilton polynomial for $S(L)$ rises
to cubic order, $S(L)=c_0\mathbf{1}+c_1\Hf+c_2\Hf^2+c_3\Hf^3$, with a
quartic characteristic equation solvable by Ferrari's
method~\cite{Ferrari1545}; the algebraic logic of
Sections~\ref{sec:su3}--\ref{sec:diag} carries over directly.

\subsection{Future Landscape}
\label{sec:future}
DUNE Phase~II targets $5\sigma$ CP sensitivity over more than $75\%$ of
$\delta_{\mathrm{CP}}$ values~\cite{DUNE2021}; HK~\cite{HK2018} adds
complementary $L/E$ coverage; JUNO will fix $\Delta m^2_{31}$ to
$0.3\%$~\cite{JUNO2022}; and IceCube-Gen2~\cite{IceCubeGen2} and
KM3NeT/ORCA will probe the ordering via Earth matter effects. The
algorithmic frontier continues to move toward more complex propagation
scenarios beyond the constant-density LBL setting of this work.

\section{Conclusion}
\label{sec:conc}

None of the individual results here is new in isolation; the value lies
in how they fit together. The starting point is the Cayley--Hamilton
structure of $S(L)$. Any analytic function of a $3\times3$ matrix is a
quadratic polynomial in it (Eq.~\ref{eq:chpoly}), so the evolution
operator in constant-density matter is fixed once the three eigenvalues
are known, and the Vandermonde system (Eq.~\ref{eq:vandermonde}) is
nothing more than the bookkeeping that makes this concrete. From there
the DMP rotation reads naturally as a resummation rather than an
approximation: it eliminates the near-degenerate eigenvalue behind the
$\Ahat\to1$ divergence and replaces the unbounded $(1-\Ahat)^{-1}$ with
the bounded $\epsilon_0\le0.015$ (Eq.~\ref{eq:eps0}). The same operator
language carries the wave-packet analysis, which puts the matter-dressed
coherence lengths in the $10^4$--$10^6\km$ range and hence
$L/L_{\mathrm{coh}}$ at the
$10^{-3}$--$10^{-2}$ level---decoherence is small, though by a more modest
margin than is sometimes assumed. Finally, the closed-form Jacobians in
the NuFIT~6.0 basis (Eqs.~\ref{eq:dP_dtheta13}--\ref{eq:dP_ddm31}) make
linearized uncertainty propagation a matter of one matrix product, with
Feldman--Cousins held in reserve for the boundaries where the Gaussian
picture breaks.

We are not proposing a replacement for NuFast-LBL. For the
$\mathcal{O}(10^6)$--$\mathcal{O}(10^7)$ evaluations a production fit
demands, the Denton--Parke algorithm~\cite{Denton2024} is the right
choice and we say so plainly. The analytic route earns its place
elsewhere: where one wants continuity in every parameter, limits that can
be read off by hand, and Jacobians that propagate correlated
uncertainties in closed form. Those are the demands of interpretation and
validation, not of likelihood minimization, and that is the work this
framework is meant for.

The extensions to non-standard interactions and sterile-neutrino
$\su(4)$ topologies follow the same algebraic logic, with the
Cayley--Hamilton polynomial rising to cubic order for the $4\times4$
case. As DUNE, HK, and JUNO approach their design sensitivities, the
interplay between matter effects, CP violation, and parameter
degeneracies will require exactly the kind of algebraic transparency this
framework is designed to provide.

\section*{Acknowledgements}
We thank the NuFIT collaboration for maintaining their publicly
accessible global fit at \href{http://www.nu-fit.org}{\texttt{nu-fit.org}},
and Peter Denton and Stephen Parke for making the NuFast-LBL code and the
\textit{Looking Glass} comparison publicly available. The experimental
work of the Super-Kamiokande, SNO, KamLAND, Daya Bay, T2K, and NOvA
collaborations underpins every quantitative statement here. We express our deepest gratitude to Professor Takaaki Kajita for his invaluable guidance and thorough reviews, which significantly shaped and refined this manuscript into its final form.

\paragraph{Funding.}
This research received no specific funding from public, commercial, or
not-for-profit agencies.

\paragraph{Data Availability.}
The NuFIT~6.0 parameters and covariance are available at
\url{http://www.nu-fit.org}. The NuFast-LBL code is available at
\url{https://github.com/PeterDenton/NuFast-LBL}.

\paragraph{Conflicts of Interest.}
The authors declare no competing interests.


\appendix

\section{Physical Constants and Reference Values}
\label{app:constants}
\begin{table}[H]
\centering
\caption{Physical constants and reference matter configuration.}
\label{tab:constants}
\begin{tabular}{ll}
\toprule
Quantity & Value \\
\midrule
$G_F$    & $1.1663787\times10^{-5}\GeV^{-2}$ \\
$m_N$    & $0.939\GeV$ \\
$1\km$   & $5.06773\times10^{9}\eV^{-1}$ \\
$\rho$   & $2.6\gcm$ (reference T2K/HK) \\
$Y_e$    & $0.5$ \\
$V_{\mathrm{CC}}$ & $9.8\times10^{-14}\eV$ \\
$E_{\Ahat=1}$ & $\approx13\GeV$ ($\rho=2.6\gcm$) \\
$E_{\mathrm{res}}$ & $\approx12\GeV$ ($\rho=2.6\gcm$) \\
\bottomrule
\end{tabular}
\end{table}

\section{The Jarlskog Invariant}
\label{app:jarlskog}
The Jarlskog invariant is the rephasing-invariant measure of CP violation
in three-generation mixing~\cite{Jarlskog1985}:
$\mathrm{Im}(U_{\alpha i}U^*_{\alpha j}U^*_{\beta i}U_{\beta j})
=J\sum_{\gamma,k}\epsilon_{\alpha\beta\gamma}\epsilon_{ijk}$ for
$\alpha\neq\beta$, $i\neq j$. In the PDG parameterization
$J=s_{12}c_{12}s_{23}c_{23}s_{13}c^2_{13}\sin\delta_{\mathrm{CP}}
=\JCP\sin\delta_{\mathrm{CP}}$; with NuFIT~6.0 values,
$\JCP\approx0.0335$ and, with $\delta_{\mathrm{CP}}\approx212^\circ$,
$|J|=|\JCP\sin\delta_{\mathrm{CP}}|\approx0.018$. In matter, $J\to\Jmat$ via
the Naumov--Harrison--Scott identity (Eq.~\ref{eq:jmat}).

\section{Gell-Mann Decomposition of the Hamiltonian}
\label{app:gellmann}
The diagonal Gell-Mann components of $\Hf$ are
\begin{align}
  h_3&=\Tr(\Hf\lambda_3)=H_{f,11}-H_{f,22},
  \label{eq:h3}\\
  h_8&=\frac{1}{\sqrt{3}}\bigl(H_{f,11}+H_{f,22}-2H_{f,33}\bigr),
  \label{eq:h8}
\end{align}
with $h_1,\ldots,h_7$ following analogously from
$h_a=\Tr(\Hf\lambda_a)$. The off-diagonal components encode flavor
transition amplitudes; $h_3$ and $h_8$ encode the diagonal energy
differences, and $|\bm{h}|$ sets the Bloch-vector precession frequency
(Eq.~\ref{eq:precession} via Section~\ref{sec:bloch}).

\section{Remaining Jacobian Components}
\label{app:jacobian}
The two phase derivatives entering Eq.~\eqref{eq:dP_ddm31} follow from
differentiating the structure factors~\eqref{eq:structfac}. Writing
$P_{\mu e}=T_{\mathrm{atm}}+T_{\mathrm{int}}+T_{\mathrm{sol}}$,
\begin{align}
  \frac{\partial\mathcal{S}_{\mathrm{a}}}{\partial\Delta}
  &=\cos[(1-\Ahat)\Delta],
  &
  \frac{\partial\mathcal{S}_{\mathrm{i}}}{\partial\Delta}
  &=\cos(\Ahat\Delta),
  \label{eq:dstruct_dDelta}\\
  \frac{\partial\mathcal{S}_{\mathrm{a}}}{\partial\Ahat}
  &=\frac{\mathcal{S}_{\mathrm{a}}-\Delta\cos[(1-\Ahat)\Delta]}{1-\Ahat},
  &
  \frac{\partial\mathcal{S}_{\mathrm{i}}}{\partial\Ahat}
  &=\frac{\Delta\cos(\Ahat\Delta)-\mathcal{S}_{\mathrm{i}}}{\Ahat},
  \label{eq:dstruct_dAhat}
\end{align}
so that
\begin{align}
  \frac{\partial P_{\mu e}}{\partial\Delta}
  &=8s^2_{23}s^2_{13}c^2_{13}\,\mathcal{S}_{\mathrm{a}}\cos[(1-\Ahat)\Delta]
   -8\alpha\JCP\sin(\Delta+\delta_{\mathrm{CP}})
     \mathcal{S}_{\mathrm{i}}\mathcal{S}_{\mathrm{a}}
  \notag\\
  &\quad
   +8\alpha\JCP\cos(\Delta+\delta_{\mathrm{CP}})
       \bigl(\mathcal{S}_{\mathrm{a}}\cos(\Ahat\Delta)
            +\mathcal{S}_{\mathrm{i}}\cos[(1-\Ahat)\Delta]\bigr)
   +8\alpha^2 c^2_{13}c^2_{23}s^2_{12}c^2_{12}\,
     \mathcal{S}_{\mathrm{i}}\cos(\Ahat\Delta),
  \label{eq:dP_dDelta}\\
  \frac{\partial P_{\mu e}}{\partial\Ahat}
  &=8s^2_{23}s^2_{13}c^2_{13}\,\mathcal{S}_{\mathrm{a}}
     \frac{\partial\mathcal{S}_{\mathrm{a}}}{\partial\Ahat}
   +8\alpha\JCP\cos(\Delta+\delta_{\mathrm{CP}})
     \Bigl(\mathcal{S}_{\mathrm{a}}
       \frac{\partial\mathcal{S}_{\mathrm{i}}}{\partial\Ahat}
      +\mathcal{S}_{\mathrm{i}}
       \frac{\partial\mathcal{S}_{\mathrm{a}}}{\partial\Ahat}\Bigr)
   +8\alpha^2 c^2_{13}c^2_{23}s^2_{12}c^2_{12}\,
     \mathcal{S}_{\mathrm{i}}
     \frac{\partial\mathcal{S}_{\mathrm{i}}}{\partial\Ahat}.
  \label{eq:dP_dAhat}
\end{align}
Together with Eqs.~\eqref{eq:dP_dtheta13}--\eqref{eq:dP_ddm21} these give
all six closed-form components of $J=\partial P_{\mu e}/\partial\bm{p}$.
Each is a smooth analytic function of the parameters, so the Jacobian is
continuous everywhere except where a structure-factor denominator
$(1-\Ahat)$ or $\Ahat$ vanishes; those points are handled by the
removable-singularity limits of Section~\ref{sec:valid}. The disappearance
derivatives $\partial P_{\mu\mu}/\partial p_k$ follow identically from
Eq.~\eqref{eq:pmumu}.

\end{document}